# Growing length scales while approaching the glass transition

Alois Loidl,[1,a] Peter Lunkenheimer,[1] and Konrad Samwer[2]

[1] Experimental Physics V, University of Augsburg, 86135 Augsburg, Germany

[2] I. Physikalisches Institut, Universität Göttingen, 37077 Göttingen, Germany

a) Author to whom correspondence should be addressed:
alois.loidl@physik.uni-augsburg.de

## ABSTRACT

Even after decades of research, the nature of the glass transition is controversially debated: Kinetic freezing phenomena, purely local strain-mediated effects, or collective phenomena including a hidden phase transition are usually discussed. In this work, we check reports on clusters of correlated particles in supercooled liquids which grow upon approaching the glass-transition. We specifically focus on dielectric studies providing approximate access to the four-point correlation function, which allows insight into the evolution of so-called "amorphous order". Existing temperature-dependent data derived from the three-point correlation function or higher-order dielectric susceptibilities are modeled using the random first-order transition theory combined with the Vogel-Fulcher-Tammann law. This ansatz enables detailed predictions concerning normalized correlation lengths or the number of correlated particles as function of temperature and fragility. It provides a good fit to the experimental data for all materials investigated, including network and molecular glass formers and polymers. We conclude that length scales and corresponding particle numbers determined by the different methods used are at least similar and closely related to those considered by the theory. Moreover, the model predictions on the fragility dependence of the number of correlated molecules are at least approximately fulfilled. Existing criticism concerning an apparent lack of criticality in the temperature dependence of experimental correlation lengths is discussed in terms of this fragility dependence: All available data are well consistent with a hidden phase transition which is best revealed for the most fragile liquids.

## I. INTRODUCTION

When avoiding crystallization upon cooling, the viscosities of supercooled liquids drastically increase and relaxation times slow down by many orders of magnitude without significant structural changes. At glassy freezing, on laboratory time scales the dynamics falls out of thermal equilibrium and the glass-transition temperature $T_g$ is canonically defined by a relaxation time $\tau$ = 100 s or alternatively by a viscosity $\eta = 10^{12}$ Pa s. Glass-transition phenomena are important for a variety of materials, including silicate glasses, polymers, molecular glasses, ionic liquids, bio-related matter, and amorphous metals, which often are of high technological relevance. Despite enormous theoretical and experimental efforts, a microscopic understanding of continuous solidification without crystallization is still missing and controversially discussed.[1,2,3,4,5] Notably, $T_g$ is based on a purely human-made definition, and it remains a mystery why this transition temperature is observed with good approximation at two-thirds of the melting temperature $T_m$,[6] a well-defined thermodynamic quantity. Over the years, different models and theoretical concepts have been developed to understand the experimentally observed dramatic increase of viscosity when approaching $T_g$ (for critical surveys see, e.g., Refs. 4,5,7. Here we consider the proposal initiated by Adam and Gibbs[8] and later theoretically worked out in the random first-order transition (RFOT) theory by Wolynes, Kirkpatrick, Lubchenko and coworkers,[9,10] guided by theoretical concepts developed for spin glasses.[11] In this theory, supercooled liquids consist of clusters or "droplets" of correlated molecules which grow on decreasing temperature. This growth, which is determined by competition of configurational entropy and surface energy, produces an increase of the apparent hindering barriers against relaxational motion: In each cluster, the relaxation of one molecule is only possible when all the remaining molecules in this cluster reorganize as well. In RFOT theory, these clusters are characterized by a growing length scale $\xi$ or, alternatively, by a growing number of correlated particles $N_{corr}$. This increasingly correlated motion is the source of the significant super-Arrhenius behavior of the viscosity or relaxation time observed in many fragile glass formers:[3] Approaching the glass transition, it drastically enhances the effective energy barriers relevant for the molecular relaxations that are related to the viscosity of the supercooled liquid. The fragility $m$, defined by the slope at $T_g$ in the Angell plot, $\log_{10}(\tau)$ vs. $T_g/T$,[12] is about 16 for strong glasses, which are characterized by pure Arrhenius behavior, and ~170 for the most fragile non-polymeric solids.[13] Fragile liquids have relatively low viscosity at the melting temperature[6] and reveal a strong increase of $\eta(T)$ or $\tau(T)$ towards the glass transition. As mentioned above, in the Adam-Gibbs scenario[8] or in the RFOT theory[14,15] this strong temperature dependence of apparent activation energies is explained in terms of growing length scales when approaching a phase transition located below the glass-transition temperature, which is hidden by the kinetic freezing occurring around $T_g$.

However, there are also alternative theoretical proposals, where glassy freezing does not rely on growing length scales. Facilitation-based theories[16,17] and Dyre's shoving-model[4] are prominent examples. Recently, the criticism on the importance of growing length scales was reinforced by Wyart and coworkers,[18,19,20] proposing that only a minor fraction of the increase of the apparent activation energies stems from correlation effects, while the main contributions result from purely local effects strongly connected with the high-frequency shear modulus. Hence, these latter authors questioned whether growing length scales control the glass transition.[18]

In the present work, we quantitatively analyze published experimental results on correlation lengths, determined from dielectric spectroscopy, using the RFOT theory combined with the Vogel-Fulcher-Tammann (VFT) equation. The latter represents a commonly used empirical description of $\tau(T)$ or $\eta(T)$ in supercooled liquids [2,6,12] and was also derived in various theoretical approaches, including the Adam-Gibbs theory.[8] Combining RFOT and VFT enables quantitative predictions concerning the absolute value and the temperature dependence of the number of correlated particles. We find reasonable agreement of experimental data and model predictions, with remarkable results concerning their temperature dependence, which strongly depends on fragility. In our analysis of experimental results, here we only consider one theoretical view of glass formation, and we do not aim to judge the validity of the mentioned competing approaches. However, we document that the RFOT theory, combined with the VFT law, reasonably well explains a variety of experimental results on growing length scales, derived by different groups, utilizing different dielectric techniques. We hope this work will trigger similar quantitative checks, based on competing theories.

## II. MODEL CALCULATIONS

As discussed above, it is intuitive to assume that growing time scales observed in glassy freezing are correlated with the growing lengths presumed in various models describing the dynamics of spin glasses.[11] Growing length scales of cooperatively coupled clusters are also predicted by RFOT theory[14,21] where the correlation length $\xi$ is related to the configurational entropy $s_c$ via[14]

$$\xi/a = \frac{5.1}{(s_c)^{2/3}} \quad . \qquad (1)$$

Here $a$ is defined as molecular length and corresponds to an approximate nearest-neighbor distance. Within this theory, $s_c$ can be expressed in terms of normalized relaxation times, $s_c = 32/\ln(\tau/\tau_0)$.[14] Assuming a VFT law, $\tau = \tau_0 \exp[DT_{VF}/(T-T_{VF})]$,[3,12] for $\tau(T)$, $s_c$ then depends on its exponent:

$$s_c = \frac{32(T-T_{VF})}{D\,T_{VF}} \qquad (2)$$

Here $\tau_0$ corresponds to the high-temperature limiting relaxation rate, often assumed to be around $10^{-14}$ s,[12] $D$ is the strength parameter,[3,12] and $T_{VF}$ the Vogel-Fulcher temperature where the relaxation times would diverge if freezing could be avoided.[3] Combining Eqs. (1) and (2) and using the approximate relations $T_{VF}/T_g = (m-16)/m$[6] and $D = 590/(m-16)$,[12] one derives:

$$(\xi/a)^3 \approx N_{corr} = N_{corr}^0 \left[ 1 + \frac{m\,(T/T_g - 1)}{16} \right]^{-2} \qquad (3)$$

where $N^0_{\mathrm{corr}} = N_{\mathrm{corr}}(T_g) \approx 176$ can be derived from Eq. (1) using $s_c(T_g)$ obtained from $s_c = 32/\ln(\tau/\tau_0)$ with $\tau(T_g)$ = 100 s. Surprisingly, it is constant for all supercooled liquids. Equation (3) provides a description of the temperature dependence of the number of correlated particles/molecular units which only depends on fragility $m$ and glass-transition temperature $T_g$ for a given material. Both are known for most glass formers and documented in literature.[12,22] Equation (3) implies a number of unexpected predictions, which all are accessible experimentally: (i) at the glass-transition temperature all supercooled liquids, independent of structural details, molecular size or bonding strength should reveal clusters with a universal number of correlated molecules $N^0_{\mathrm{corr}} \approx 176$; ii) the temperature dependence of $N_{\mathrm{corr}}$ is a function of glass-transition temperature and fragility only; (iii) the use of the VFT law automatically guarantees that $N_{\mathrm{corr}}$ diverges at $T_{\mathrm{VF}}$ and, hence, reveals true criticality, and iv) even the strongest glass formers with $m$ = 16 are characterized by significant temperature dependence $N_{corr} = N^0_{corr}(T_g/T)^2$. The latter seems counterintuitive as for $m = 16$ the relaxation time follows an Arrhenius law suggesting temperature-independent energy barriers. However, as mentioned by Berthier et al.,[23] even strong glasses possibly develop non-trivial dynamic correlations.

It seems interesting and can serve as additional plausibility check to discuss the fragility dependence of the correlation lengths at melting assuming the validity of the two-thirds rule $T_g \approx 2/3\ T_m$.[6] According to Eq. (3), strong liquids with $m$ = 16 are characterized by a number of correlated molecules $N_{corr}(T_m) \approx 78$, while fragile liquids with $m$ = 170 reveal a drastically smaller number, $N_{corr}(T_m) \approx 4$. Hence, in the framework of RFOT theory the correlation number mirrors the experimentally observed strong fragility dependence of the viscosity of liquids close to $T_m$.[6] In addition, at the upper boundary of fragility for non-polymeric solids[13] the normalized correlation length $\xi/a \approx 1.6$ just exceeds the typical distance to the next-nearest neighbor shell. Low-viscosity liquids with $m$ close to 170 are characterized by small correlation lengths at melting, and seemingly in these melts crystallization is favored compared to supercooling. The above considerations also make it obvious that, against common belief, fragile liquids should not feature stronger particle correlations or cooperativity than strong ones. Instead, within the RFOT framework, it is solely the strong temperature dependence of these correlations that characterizes a fragile glass former.

## III. LENGTH SCALES MEASURED BY DIFFERENT EXPERIMENTAL METHODS

In RFOT theory, supercooled liquids are described as mosaic of droplets, which are termed "correlated volumes",[10] "cooperative regions",[14,21] or "rearranging units".[21] Their temperature-dependent size is denoted "coherence",[10] "correlation",[10,14,21] or "cooperativity length".[14,15] Thus, these objects remind of the cooperatively rearranging regions (CRRs) as proposed by Adam and Gibbs,[8] although it was pointed out in Ref. 21 that their size should be larger and reveal stronger temperature dependence that predicted by that older theory. The droplets assumed within the RFOT theory imply a non-uniform, heterogeneous nature of supercooled liquids. Indeed, it was noted that "dynamic heterogeneities" are "intrinsic to the mosaic picture of the deeply supercooled liquid".[14] Notably, dynamic heterogeneity is a well-known concept in glass physics since long: It was shown to explain one of the hallmark features of supercooled liquids,[4] namely their non-exponential relaxation dynamics, which is nowadays commonly ascribed to a distribution of relaxation times caused by heterogeneity.[24,25] Non-exponentiality, which can be quantified, e.g., by the stretching exponent $\beta$, is usually found to become stronger upon cooling.[26,27] This may suggest an increase of the length scales associated with this heterogeneity when approaching the glass transition, reminding of the predicted growth of the droplet sizes invoked within RFOT theory to explain another typical feature of supercooled liquids – the mentioned non-Arrhenius behavior. In recent literature, there is no clear consensus concerning the terms cooperativity or heterogeneity and they are often used almost synonymously, despite in principle describing quite different concepts. While some interrelations may exist, it was pointed out in Ref. 28 that cooperativity seems to be less well defined.

Various experimental methods were used to detect length scales in supercooled liquids. However, in some cases their exact nature remains rather unclear. It is well accepted that multidimensional NMR, revealing values of ~ 3 nm, detects heterogeneity, related to relaxation-time distributions in supercooled liquids.[29,30] In contrast, molecular cooperativity lengths ranging from 1.4 – 3.4 nm were estimated by analyzing the spectral shape of the boson peak, but their values were directly compared to the heterogeneity lengths from NMR.[31] Donth and coworkers[32,33] determined lengths between 1.0 – 3.6 nm from heat-capacity measurements, identifying them with the sizes of CRRs, but the justification for this assignment was doubted in Ref. 28. Recently, many of these results were summarized and interpreted as "average correlation length of the CRR" by Schawe *et al.*[34] They found that, at $T_g$, all these lengths for various classes of glass formers seem to fall into a narrow range of 2 – 4 nm. Notably, this finding is in accord with the RFOT theory prediction of a universal relative correlation length $\xi(T_g)/a \approx \sqrt[3]{N_{\mathrm{corr}}^{0}} \approx 5.6$ [see Eq. (3)]. Assuming a correlation length of 3 nm results in a nearest-neighbor distance of ~ 0.5 nm, which is compatible with the observation of the first-sharp diffraction peak in many supercooled liquids (see, e.g., Ref. 35).

Next, we turn to the determination of temperature dependent correlation lengths from dielectric spectroscopy, which is the main scope of this work. Theoretical concepts developed in the field of spin glasses[11] evidenced that conventional two-point correlation functions are blind to long-range "amorphous order"[5,23,36] suggested by RFOT and that the respective lengths can only be detected via the four-point correlation function $\chi_4$.[23] For supercooled liquids, the importance of four-point correlators, simultaneously probing correlations at two points and two times, and, hence, measuring correlations of fluctuations, was established by Parisi and others,[37,38] Bennemann *et al.*[39] and Berthier *et al.*[40,41] Experimentally, four-point correlation functions are hardly accessible, except in cases one can monitor trajectories of individual particles. However, it was shown that they can be approximately measured via non-linear dielectric spectroscopy.[42] Moreover, a lower bound can be obtained by evaluating the three-point correlation function $\chi_T$, which is defined via the temperature derivative of the time-dependent two-point correlation function and, making some assumptions, can be estimated from derivatives of the linear susceptibility or relaxation time.[23,28,41] The relation between $\chi_4$ and $\chi_T$ is not trivial, and here we follow arguments provided in Refs. 28,43,44 and analyze the number of correlated particles assuming the relation $\chi_4 \sim \chi_T$.

Several measurements of $\chi_T$ were performed for different classes of materials Refs. 28,41,45,46,47, providing detailed results on the temperature dependence of $N_{corr}$. Notably, the authors of Refs. 45,46,47 analyzed their results assuming $\chi_4 \sim \chi_T^2$. Hence, as explained above, in fitting their results we used the square root of their published $N_{corr}$ values, to stay consistent with the analysis of the remaining data. In the following years, triggered by the proposal that $N_{corr}$ can be detected via non-linear dielectric measurements,[42] numerous such experiments on disordered materials were performed Refs. 36,44,48,49,50,51,52,53,54,55,56. This includes aging experiments,[48] studies of plastic crystals,[50,51,55] non-linear measurements under hydrostatic pressure,[52] and even experiments of 5$^{th}$ order susceptibilities.[53,54,56]

In the present work, we do not aim to clarify the mentioned ambiguities about the true nature of the length scales detected by different methods and their relation to the correlation length $\xi$ (or number of correlated particles $N_{corr}$) considered within the RFOT theory. Instead, we pragmatically provide a critical comparison of all these published results to the expectations from the combination of the RFOT theory with the VFT law, promoted in section II, and fit a representative selection utilizing Eq. (3).

# IV. COMPARISON OF PREDICTIONS OF RFOT THEORY WITH EXPERIMENTAL RESULTS

For a detailed comparison of results from different dielectric experiments and theoretical predictions we chose propylene carbonate (PC) and glycerol, two canonical molecular glass formers, with glass transition temperatures $T_g$ = 159 and 185 K[57] and fragilities $m$ = 104 and 53,[12] respectively. The relaxation dynamics of both compounds is well characterized by broadband dielectric spectroscopy covering more than 17 decades in frequency.[58] In both materials, the VFT law specifically well describes the temperature dependence of relaxation times close to $T_g$, which seems important for the present analysis. In Fig. 1 we collected the existing dielectric results on the temperature dependence of the number of correlated molecules. We present the $N_{corr}(T)$ traces normalized to unity at the glass-transition temperature, as RFOT theory predicts a universal $N_{corr}^0 \approx 176$. We normalized the results as the absolute values may be rather uncertain, containing a number of approximations.[14] For both compounds, Figure 1 compares $N_{corr}(T)$ from third- and fifth-order susceptibilities[53] with results from three-point susceptibilities[28,45,59] as indicated in the figure legend. For the fifth-order susceptibility $\chi_5$, we assumed spherical compact clusters, leading to the relation of $\chi_5 \sim N_{corr}^2$.[53,56] As mentioned above, for the three-point susceptibilities determined by Fragiadakis *et al.*[45,47] we used $N_{corr} \sim \chi_T$ [28,43,44] and not $\sim \chi_T^2$ proposed in that work.

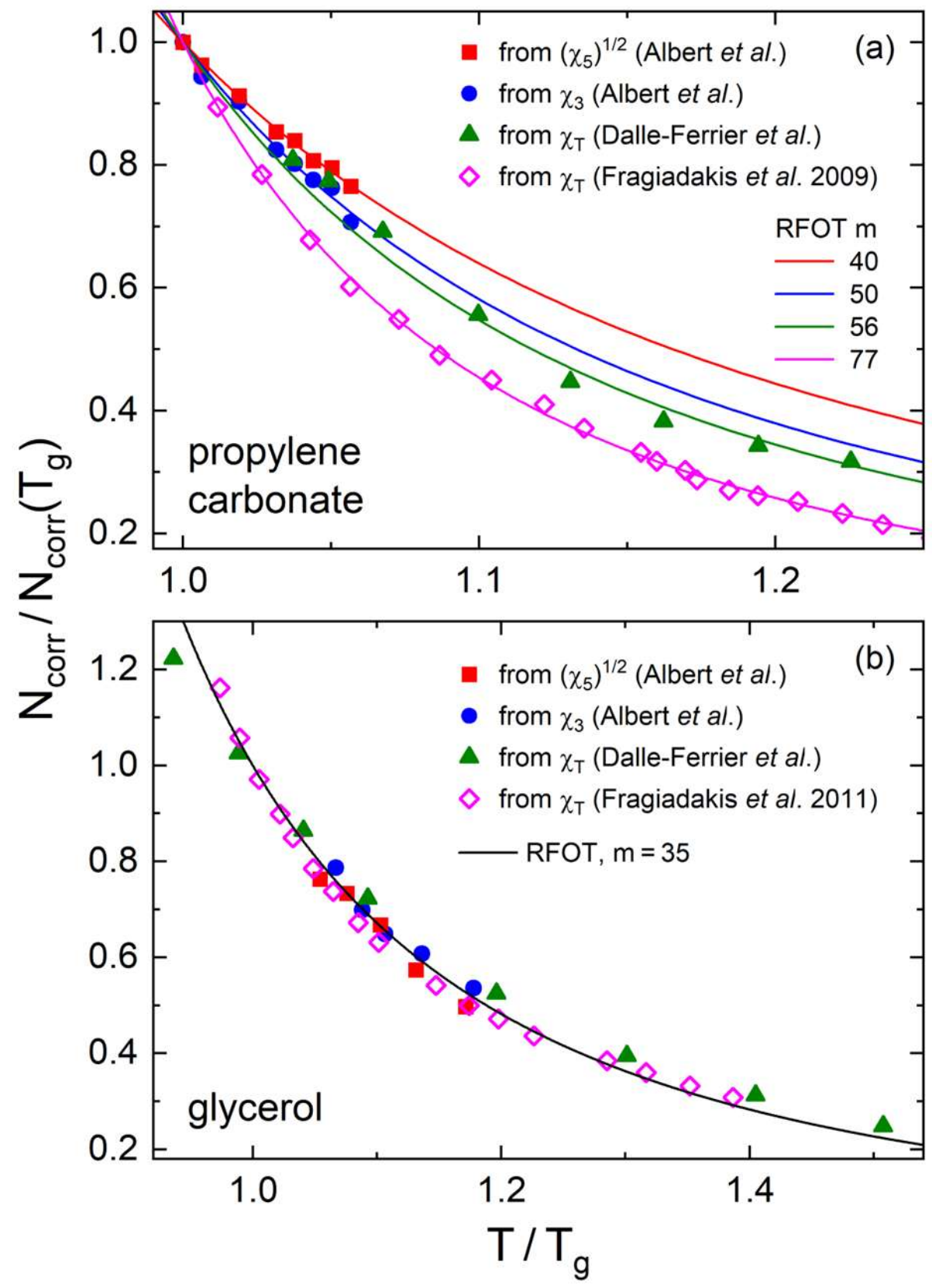

FIG 1: Normalized number of correlated molecules, $N_{\mathrm{corr}}/N_{\mathrm{corr}}(T_g)$, *vs.* the $T_g$-normalized temperature: (a) propylene carbonate and (b) glycerol as derived from different groups, utilizing different dielectric techniques as indicated in the legend: fifth order ($\chi_5$) (red squares) and third order ($\chi_3$) (blue spheres) non-linear susceptibilities from Albert *et al.*,[53] three-point correlation function $\chi_T$ from Dalle-Ferrier *et al.*[28] (green triangles) and Fragiadakis *et al.*[45,47] (magenta rhombs). The lines represent fits using Eq. (3), combining the RFOT theory with the VFT law, with fragility *m* as the only free parameter. In PC, we fitted the different measurement results separately, and in the case of $\chi_T$ from Ref. 45 we restricted the fit to the shown temperature regime. In glycerol, revealing only minor variance of the different experimental results, we fitted all data sets by a single fragility value. The two data points farthest below $T_g$, which may be measured out of equilibrium, were ignored. For each fit, the *m* values are indicated in the figure.

All the results documented in Fig. 1 were fitted using Eq. (3) with fragility *m* as the only free parameter. Of course, it would be striking if the *m* values derived from these fits would coincide with those determined analyzing the relaxation dynamics. Apparently, this is not the case: In PC [Fig. 1(a)] the different experimental techniques result in some spread of the obtained temperature dependences of $N_{\mathrm{corr}}$. First, all the determined fragilities as indicated in the figure are significantly too low, compared with the value $m = 104$ reported from relaxation data.[12] In addition, the three-point susceptibilities reveal larger fragilities than those derived from non-linear dielectric results. In PC, Fragiadakis *et al.*[45] reported $N_{\mathrm{corr}}$ values up to temperatures extending beyond $2T_g$. A fit to the complete temperature range results in $m = 81$, only slightly enhanced compared to the result shown in Fig. 1(a) ($m = 77$). We restricted the fit to low temperatures, as at higher temperatures a crossover to a power-law behavior may be expected.[28] In glycerol ($m = 54$),[12] the results are more convincing, and the temperature dependent number of correlated molecules can be well described by a unique fragility, which, however, is by a factor ~ 1.5 too low. Nevertheless, the glycerol data provide a convincing test that the different dielectric techniques probe the same observable and, in addition, that the RFOT theory provides a good description of the experimentally observed temperature dependence of the number of correlated particles.

In Fig. 2, we check for a possible fragility dependence of $N_{\mathrm{corr}}(T)$, which is predicted by RFOT theory [see Eq. (3)]. We selected three sets of measurements of glass forming compounds spanning the range from strong to fragile and analyzed results on $B_2O_3$,[28] 3-fluoro aniline (FAN)[49] and decahydro-naphthalene (decaline).[28] $B_2O_3$ is a covalent-network-forming borate, a relatively strong glass former with glass-transition temperature $T_g = 554$ K and fragility $m =$ 32.[12] FAN, consisting of small rigid molecules, is rather fragile with a glass transition at 172 K[60] and fragility $m$ = 89, which we determined from the VFT fit of $\tau(T)$ shown in Ref. 49. Finally, decaline is an extremely fragile molecular glass former with $T_g$ = 135 K and $m$ = 147.[61] $N_{corr}(T)$ of $B_2O_3$ and decaline was determined by the three-point correlation function,[28] while for FAN it was deduced from third-order nonlinear dielectric spectroscopy.[49]

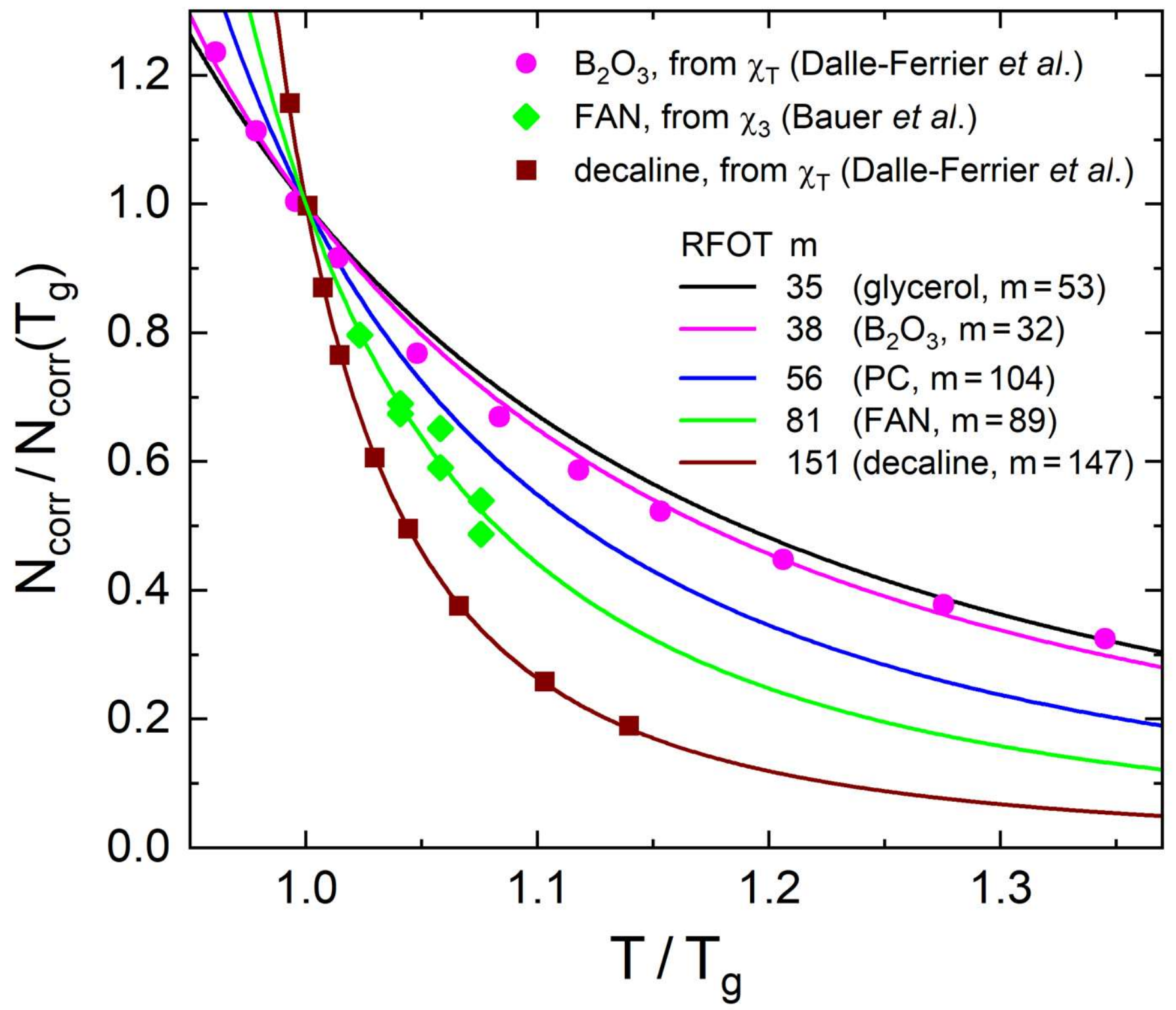

FIG. 2: Normalized number of correlated particles, $N_{corr}/N_{corr}(T_g)$, versus $T/T_g$ as reported for $B_2O_3$,[28] 3-fluoro aniline (FAN),[49] and decaline.[28] The dielectric data of Dalle-Ferrier *et al.*[28] were derived via the analysis of the derivative of the two-point correlation function, while those by Bauer *et al.*[49] represent third-order non-linear results. The solid lines show fits of the experimental data by Eq. (3) with fragility as the only free parameter. Fragilities $m$ obtained from the fits and from relaxation data are indicated in the figure legend. In addition, we added the $m$-averaged fit curve of PC [Fig. 1(a)] and the global fit of glycerol shown in Fig. 1(b).

The temperature dependences of the normalized number of correlated molecules of these three glass formers were fitted using Eq. (3), again with fragility $m$ being the only free parameter. The fits are shown by the solid lines in Fig. 2, and the resulting $m$ values are indicated in the figure legend. For these materials, fragilities derived by fitting the number of correlated molecules and those deduced from the relaxation dynamics (indicated in brackets) compare reasonably well. For comparison, we also included the results of the fits for PC and glycerol, documented in Fig. 1. For PC we plot an $m$-averaged curve using $m$ = 56 and for glycerol we show the global fit to all data with $m$ = 35. As mentioned earlier, both fragilities are significantly too low compared to those determined from the relaxation dynamics. This also becomes apparent in Fig. 2 by the sequence of the $N_{corr}(T)$ results: Glycerol being an intermediate glass former exhibits the weakest temperature dependence, although $B_2O_3$ is reported to have smaller fragility.[12] In the figure, PC is located at intermediate fragilities. However, usually it is characterized as fragile glass former ($m$ = 104)[12] and, in Fig. 2, thus should be located well below FAN. Concerning PC, one must mention two points: i) The experimental results from different sources provided in Fig. 1 reveal considerable deviations and ii) PC is the only conventional glass former that strongly deviates from the overall correlation of the Böhmer plot, *m vs.* $\beta$.[12] It exhibits a stretching exponent $\beta(T_g) = 0.73$, which according to the general correlation of supercooled liquids, $m = 250(\pm 30) - 320\,\beta$,[12] would characterize PC as significantly stronger glass former with $m \leq 46$.

## V. CONCLUDING REMARKS

In this work, we have presented a critical review of published results on correlation lengths in various supercooled liquids approaching the glass transition. We have compared the experimental data with model predictions derived by a combination of the RFOT theory with the VFT law and overall found remarkable agreement: i) The deduced prediction concerning the temperature-dependence of the relative number of correlated molecules [Eq. (3)] is well fulfilled in all investigated glass formers. This indicates that the length scales and corresponding particle numbers determined by the vastly different methods employed are all somehow related to those considered in the RFOT theory. Concerning the theory's expectation of a universal $N_{corr}$ (or $\xi$) at $T_g$, we want to point out that in literature there are several hints at its approximate validity.[34,41] ii) The prediction that the (relative) temperature-dependence of $\xi$ or $N_{corr}$ should only be determined by the fragility [Eq. (3)] is at least approximately fulfilled. Figure 2 provides the expected and rather significant correlation of $N_{corr}(T)$ with fragility, except for PC. iii) The general criticism concerning the importance of growing length scales (or correlated particle numbers) when approaching glassy freezing, namely, that all their observed temperature dependences are far from criticality, is not convincing: While the analyzed $N_{corr}(T)$ data indeed reveal only a limited relative increase when approaching the glass transition (up to a factor of five) they can be well fitted by Eq. (3) which involves a divergence at the Vogel-Fulcher temperature. Therefore, all experimental correlation numbers and lengths are well consistent with a divergence around the Vogel-Fulcher temperature, roughly corresponding to the Kauzmann temperature. Clear traces of criticality can only be observed in very fragile liquids [see $N_{corr}(T)$ of decaline in Fig. 2]. This supports the observation that for fragile liquids the Prigogine-Defay ratio approaches unity [62], indicating the closeness of a thermodynamic phase-transition temperature to $T_g$.

## AUTHOR DECLARATION

### Conflict of interest

The authors have no conflict to disclose.

### Author contributions

**Alois Loidl:** Writing – original draft (lead); Writing – review & editing (equal) Conceptualization (lead); Formal analysis (equal); Validation (equal); Visualization (supporting). **Peter Lunkenheimer:** Writing - review & editing (equal); Conceptualization (supporting); Formal analysis (equal); Validation (equal); Visualization (lead). **Konrad Samwer:** Writing - review & editing (supporting); Conceptualization (equal); Formal analysis (supporting); Validation (supporting).

## DATA AVAILABILITY

All data that support the findings of this study are available within this article.